\documentclass{article}
\usepackage{epsfig}
\usepackage{amssymb,amsmath,amsfonts}
\usepackage[spanish,english]{babel}
\usepackage[latin1]{inputenc}
\usepackage{graphicx}
\usepackage{natbib}
\usepackage{caption}
\usepackage{wasysym}
\usepackage{tikz}
\usepackage{hyperref}
\usepackage{upgreek}
\usepackage{cite}
\usepackage{transparent}
\usepackage{eso-pic}

\hypersetup{
	colorlinks=true,
	linkcolor=red, 
	urlcolor=orange,
	citecolor=blue,
	bookmarks=true,
}

\input ibvs2.sty

\begin{document}
\IBVShead{5xxx}{16 February 2019}
	
\IBVStitle{New Double Periodic Variable stars in the ASAS-SN Catalog}

\IBVSauth{ROSALES, J. A.$^1$, MENNICKENT, R. E.$^{1}$}
	
\IBVSinst{Astronomy Department, University of Concepci\'on, Concepci\'on, Chile. e-mail: jrosales@astro-udec.cl}\\

\vskip 1cm
\section*{Abstract}
We report the discovery of 3 new Double Periodic Variables based on the analysis of ASAS-SN light curves: GSD J11630570-510306, V593 Sco and TYC 6939-678-1. These systems have orbital periods between 10 and 20 days and long cycles between 300 and 600 days.

\section*{Introduction}
The Double Periodic Variable stars constitute enigmatic stellar systems discovered just in recent years. These systems are close binary stars of intermediate mass, the majority of the studied DPV are in a semi-detached stage undergoing mass transfer, and show a second photometric variability. This variation was observed for the first time in the Magellanic Clouds by \citet{2003A&A...399L..47M},  and these long periods are on average 33 times longer than the orbital period \citep{2010AcA....60..179P,2017SerAJ.194....1M,2018MNRAS.476.3039R}. To date, it has been observed that the more evolved star is generally of the A/F/G spectral type, while the companion is always of B spectral type surrounded by an optically and geometrically thick accretion disk \citep{2013A&A...552A..63B,2015MNRAS.448.1137M,2018MNRAS.476.3039R}. In addition, the second period was associated to cycles of magnetic dynamo in the more evolved star (donor), based on the Applegate mechanism \citep{1987ApJ...322L..99A} as proposed by \citet{2017A&A...602A.109S}. However, recently some changes have been observed in some light curves of DPVs and these could be related to variations in the disc size/temperature and the spot temperature/position \citep{2018MNRAS.477L..11G}. Some one of the DPVs were recently discovered using online catalogs such as ASAS-3 and most of the DPVs which have been discovered are Algol type eclipsing (DPV/E), Ellipsoidal (DPV/ELL) binaries and even a DPV of semi-regular amplitude has been found \citep{2018IBVS.6248....1R}. Therefore, we believe these catalogs are a big repository to search for new DPVs and must be reviewed periodically.

\section*{Photometric analysis and ephemeris}
We have carried out a visual inspection to find new DPVs using the ASAS-SN Variable Stars Database \footnote{\url{https://asas-sn.osu.edu/variables}} \citep{2014ApJ...788...48S,2017PASP..129j4502K,2019MNRAS.486.1907J} considering orbital periods between 10 to 40 days. We checked a total of 894 eclipsing binaries, such as the Detached Algol (EA), Beta Lyrae (EB) and Ellipsoidal (ELL) type binaries, and we found 3 new DPVs characterized by a deep primary eclipse. The orbital periods were determined using the Period Dispersion Minimization (PDM) task of IRAF\footnote{IRAF is distributed by the National Optical Astronomy Observatories, which are operated by the Association of Universities for Research in Astronomy, Inc., under cooperative agreement with the National Science Foundation.} software \citep{1978ApJ...224..953S}. The errors were estimated through visual inspection of the light curves phased with trial periods close to the minimum of the periodogram until the light curves began to increase their dispersion. Through a code written by Zbigniew Ko\l{}aczkowski specially developed for the Double Periodic Variables stars, we have disentangled the two main photometric frequencies of every system. Specifically, the code adjusts the orbital signal to Fourier series consisting of the fundamental frequency plus their harmonics. This removes the signal from the original time series letting the long periodicity in a residual light curve. As a result, we obtained the light curves without the additional frequencies in isolated light curves.

\begin{figure}[h!]
	\begin{center}
		\includegraphics[width=6cm,angle=0]{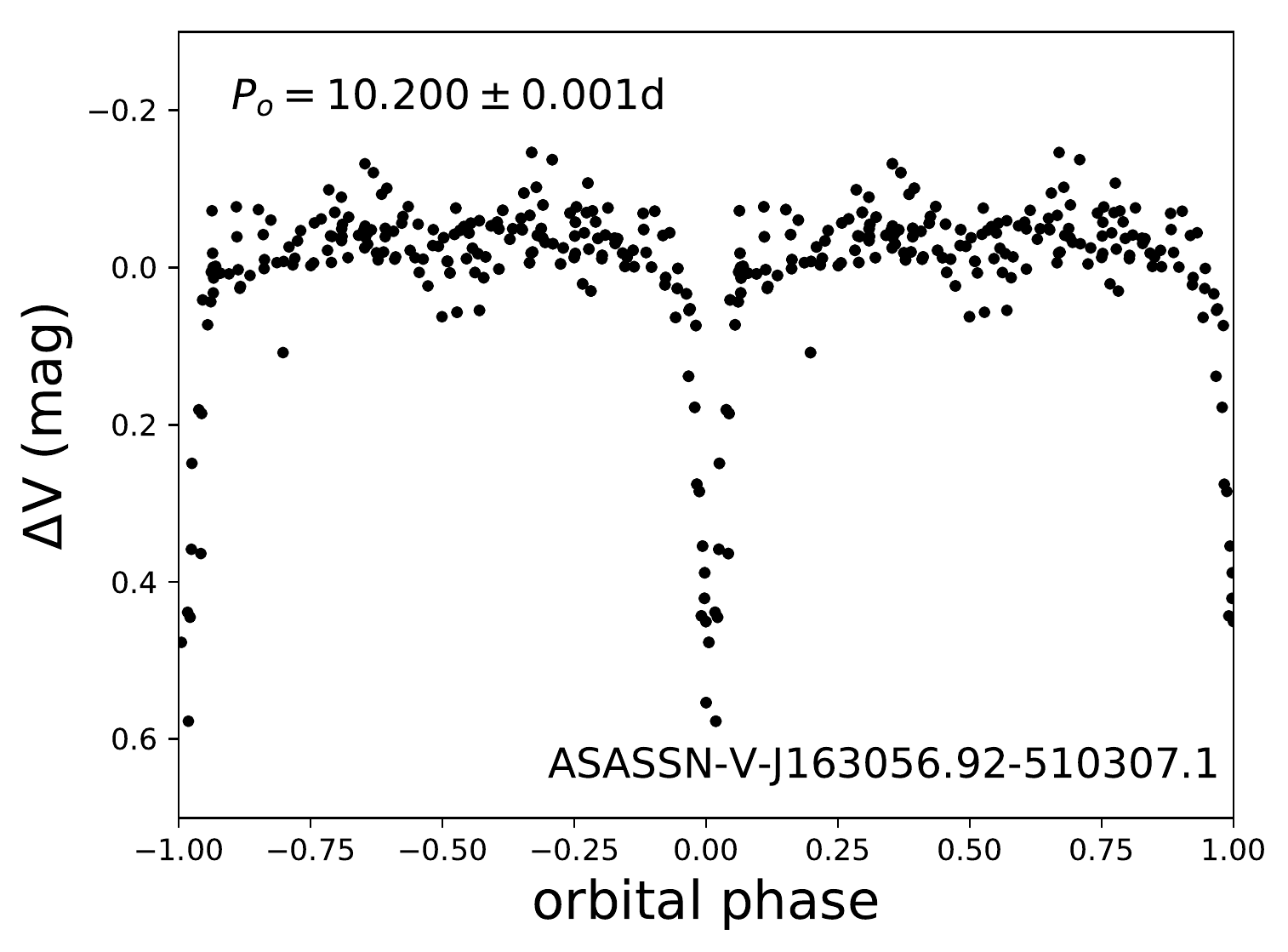}
		\includegraphics[width=6cm,angle=0]{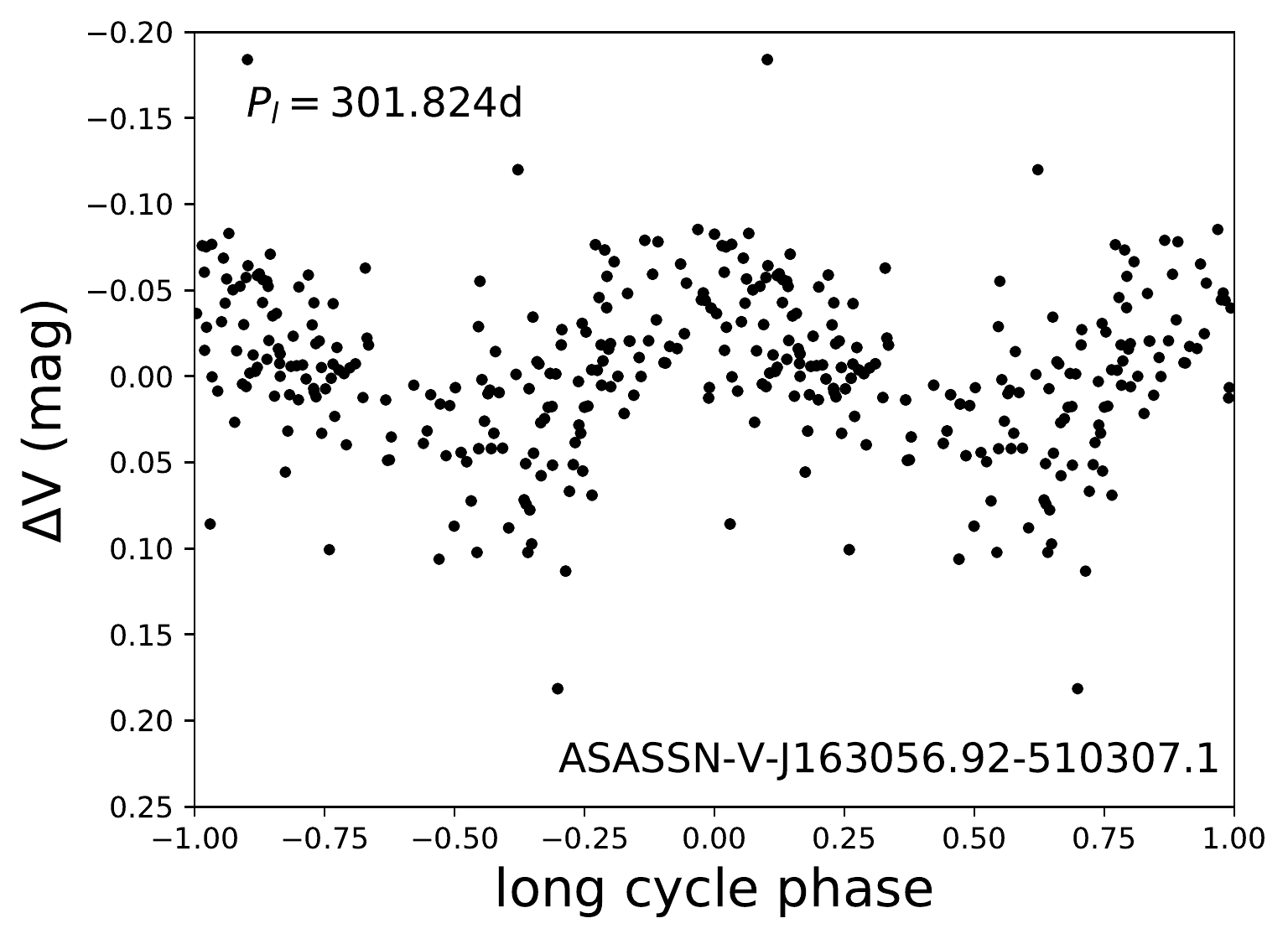}
		\includegraphics[width=6cm,angle=0]{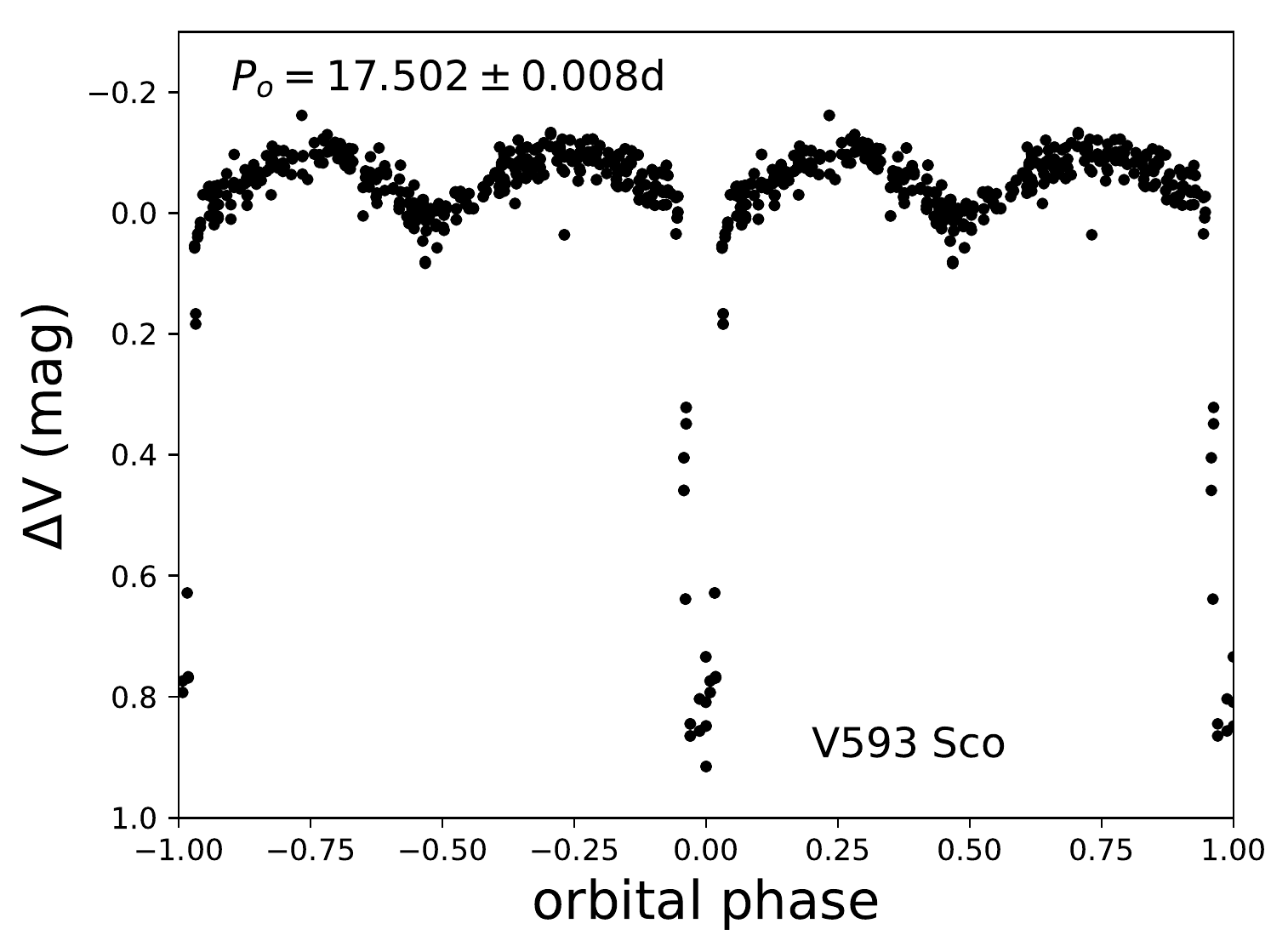}
		\includegraphics[width=6cm,angle=0]{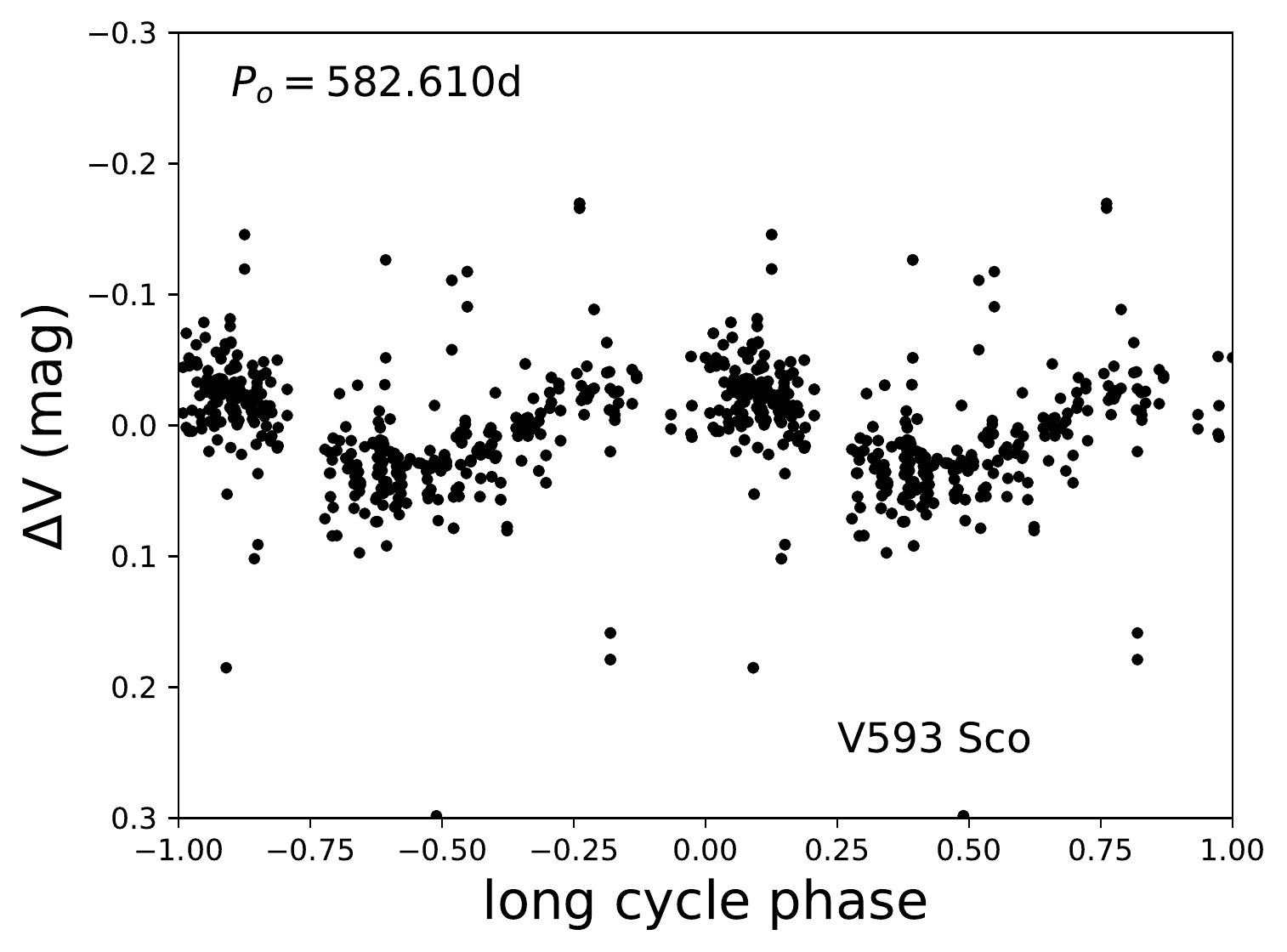}
		\includegraphics[width=6cm,angle=0]{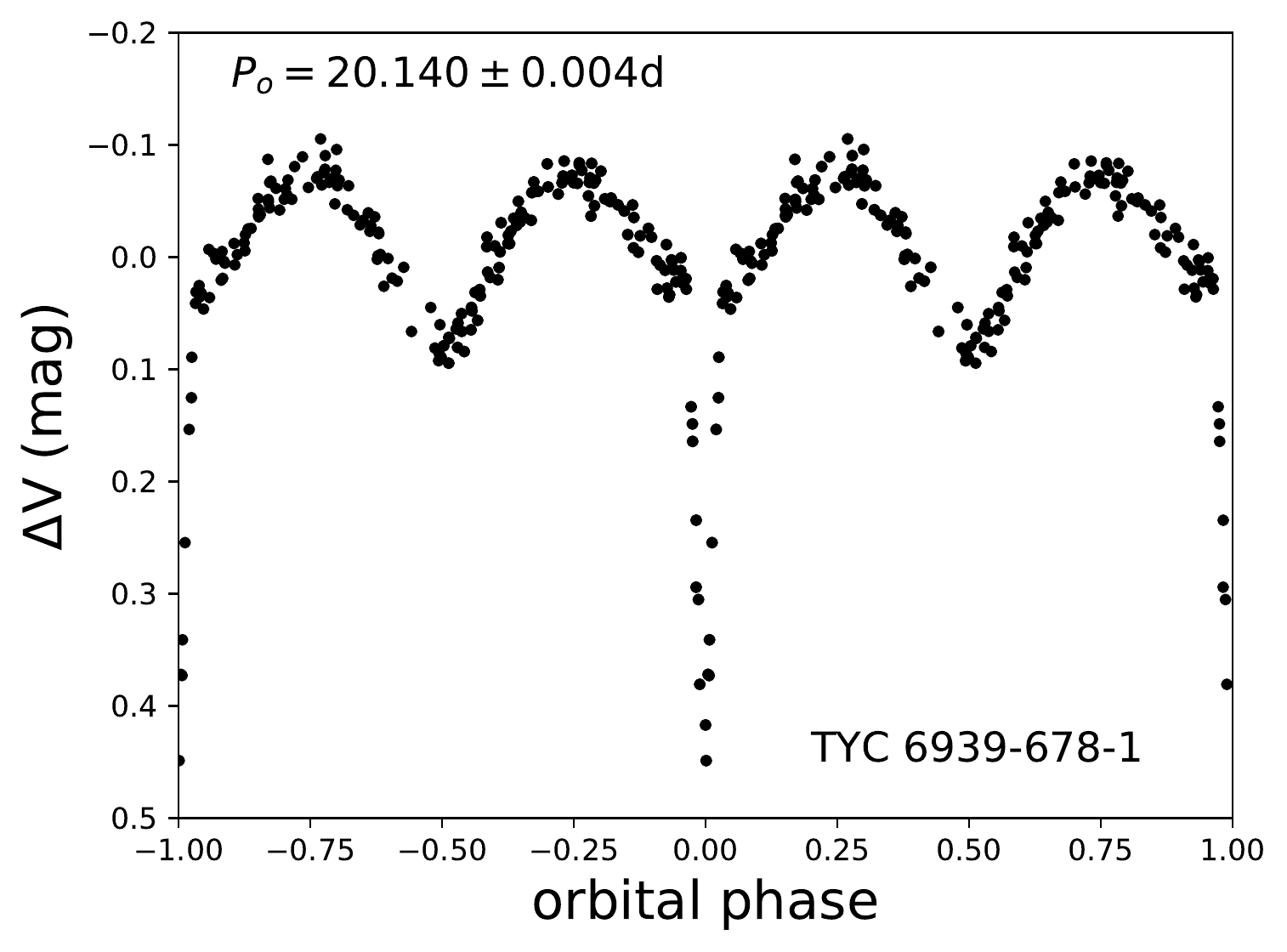}
		\includegraphics[width=6cm,angle=0]{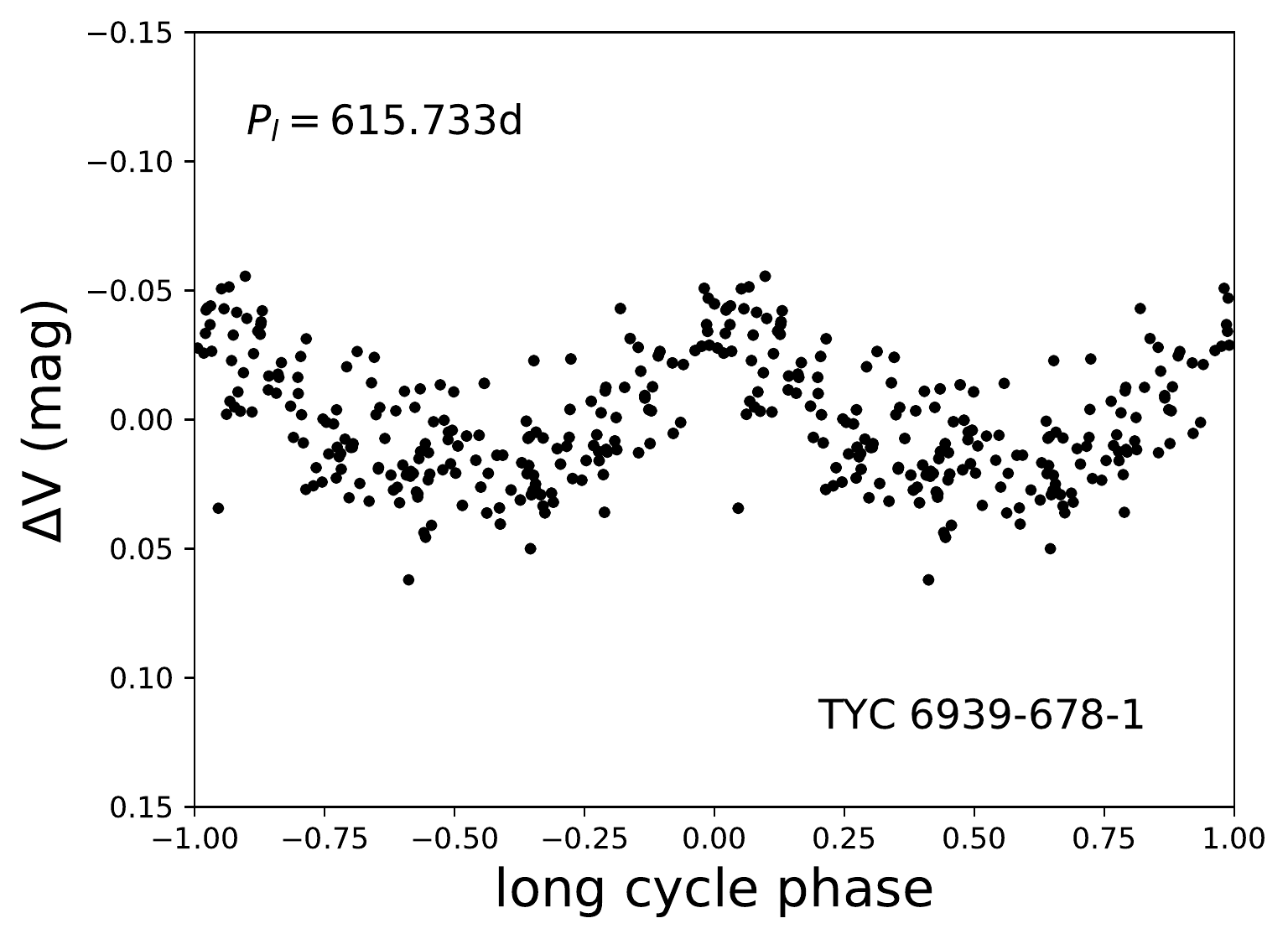}
	\end{center}
	\caption{Disentangled ASAS-SN V-band light curves of three new Double Periodic Variables stars. The left hand side corresponds to light curves phased using the orbital periods, while on the right hand side is observed the long cycle of every system.}
	\label{Fig. 1}
\end{figure}

\begin{figure}[h!]
\begin{center}
	\includegraphics[width=5.2cm,angle=0]{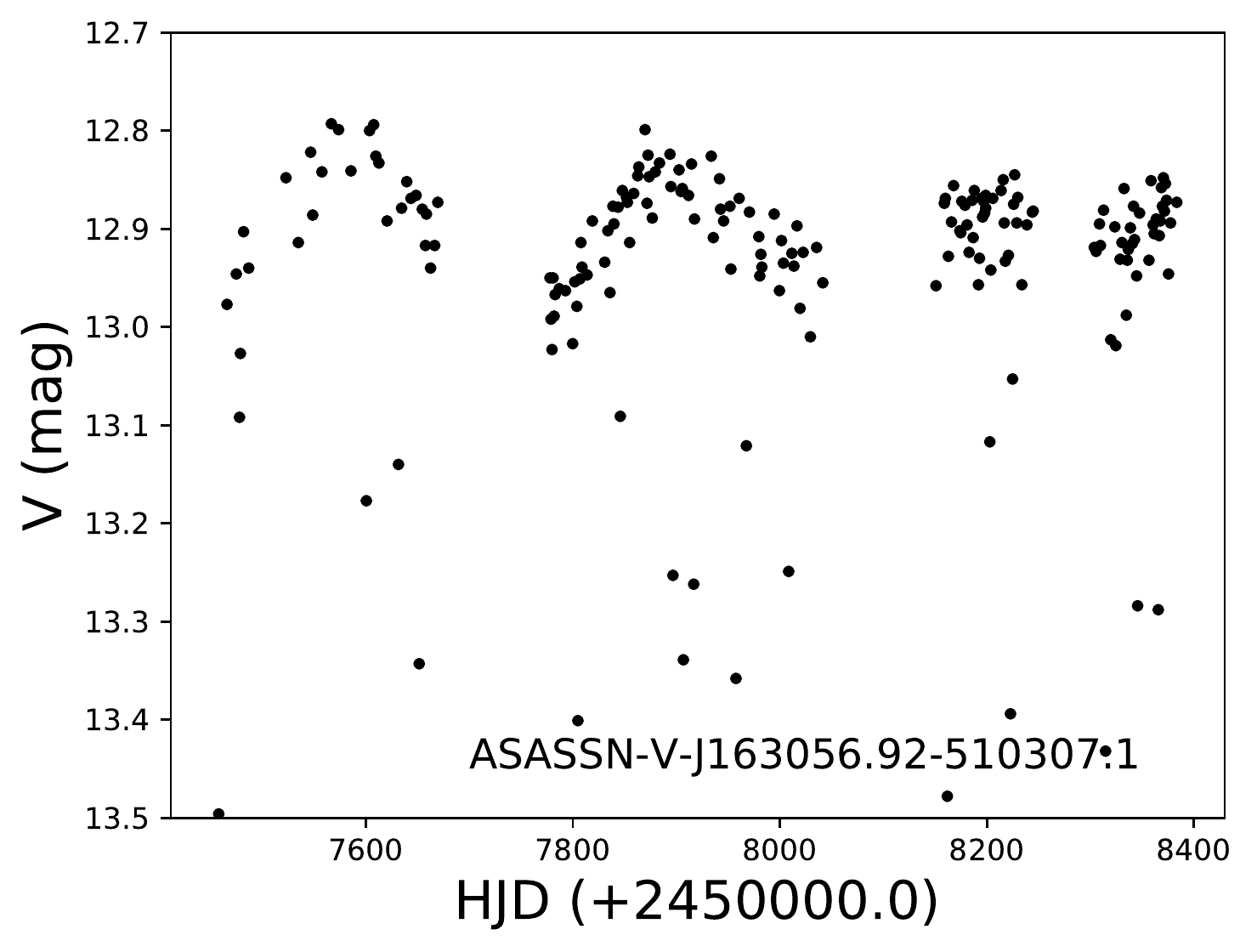}
	\includegraphics[width=5.2cm,angle=0]{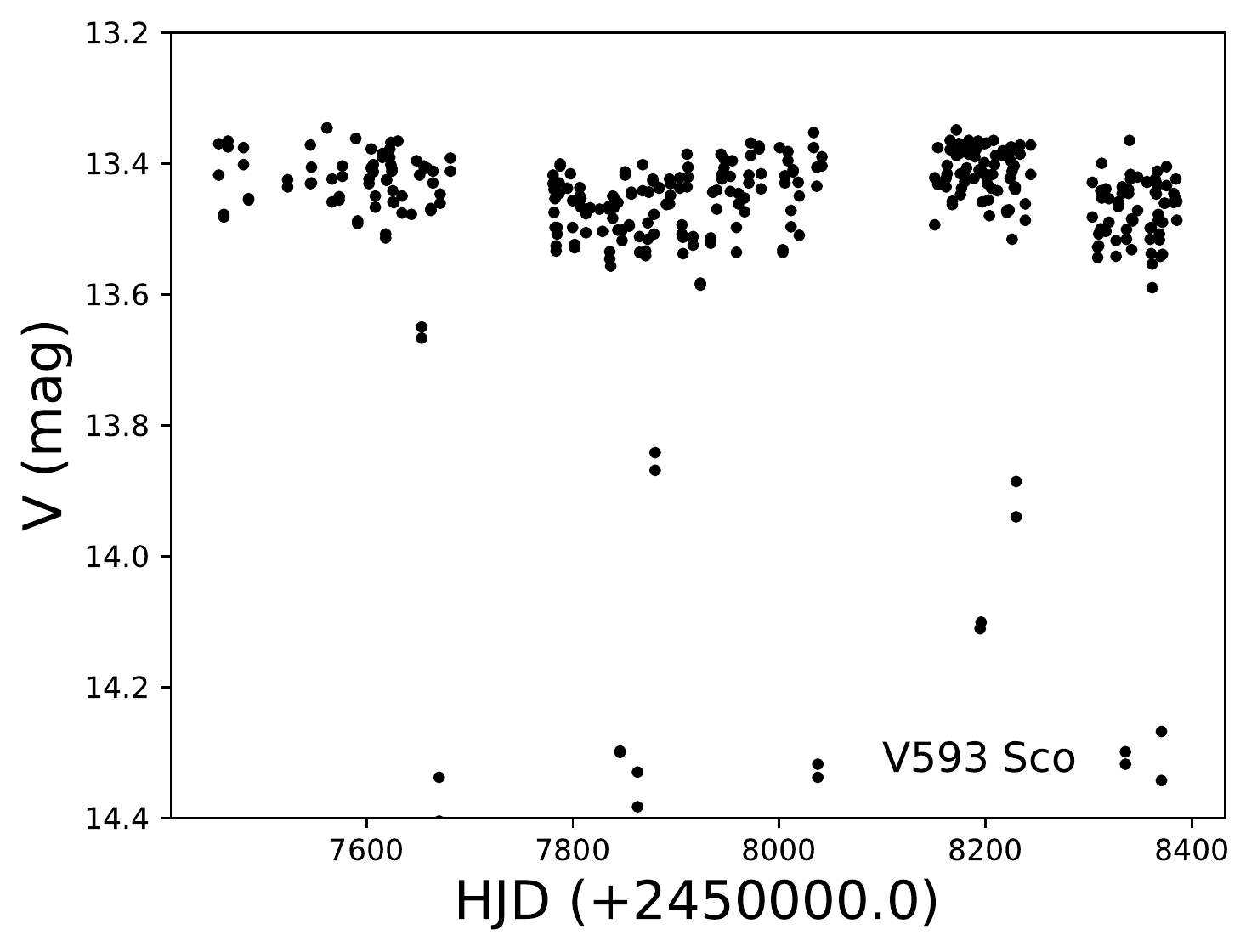}
	\includegraphics[width=5.2cm,angle=0]{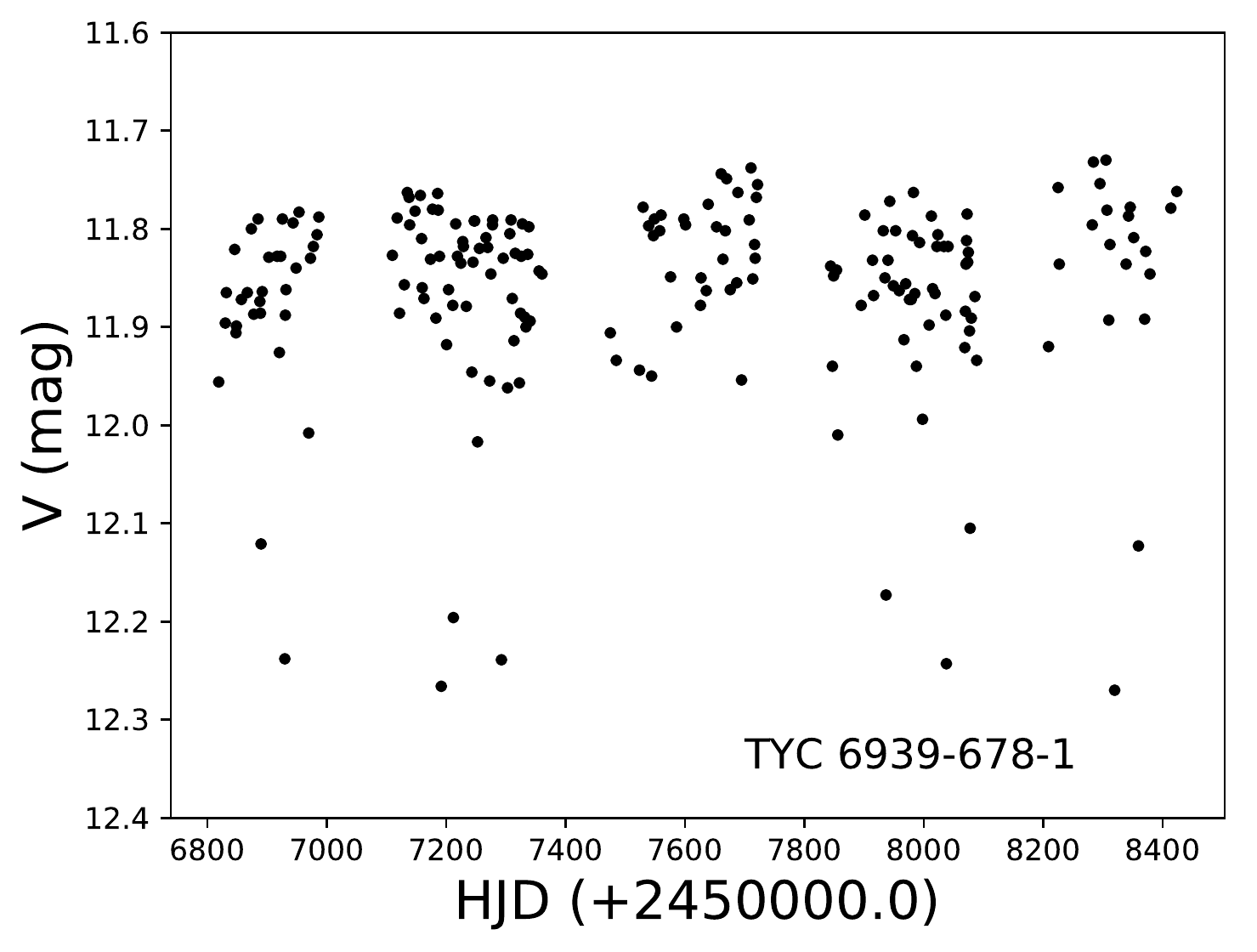}
\end{center}
\caption{Photometric data as function of the Heliocentric Julian Days wherein is easily observed the second photometric variability of the new discovered DPVs.}
\label{Fig. 2}
\end{figure}

We summarized the results in Table \ref{Table 1} and the disentangled light curves are shown in Figure \ref{Fig. 1}. They show deep primary eclipses and relatively long orbital periods. The Double Periodic Variable ASAS-SN-V J163056.92-510307.1 appears cataloged as an eclipsing Algol (EA) type in the ASAS-SN Catalog with a 0.71 mag deep primary eclipse. Apparently it is a system of low inclination, which would allow to perform a detailed study of the more evolved star and to obtain relevant information about the stellar dynamo. In addition, the full amplitude of the long cycle in the V-band is $27\%$ with respect to the total brightness of the light curve, and the long period is $P_{l}= 29.5P_{o}$. The DPV V593 Sco is other eclipsing Algol with a 1.1 mag mag deep primary eclipse, wherein the second variability is observed at the photometric data as function of the Heliocentric Julian Days (HJD, see Fig. \ref{Fig. 2}) and reveals an orbital modulation typical of a DPV of circular orbit with a full amplitude of the long cycle of $43\%$ of the total brightness. Its long period is $33$ times the orbital period, i.e. $P_{l} = 33 P_{o}$. The DPV TYC 6939-678-1 is cataloged as an eclipsing Beta Lyrae type (EB) with a 0.55 mag deep primary eclipse, within which the second photometric variability in the photometric data as a function of the HJD is easily observed, and its full amplitude is 21\% of the total brightness of the light curve as shown. In addition it shows an increase of the data dispersion around $\phi_{l}=$-0.5 and 0.5. Its respective long period is $31$ times the orbital period. A peculiarity of the long cycles in these DPVs that have been discovered is that they are characterized by a quasi-sinusoidal variability. 

Owing to the relevance of the mass loss/transfer process in close binary systems it was necessary to analyze every system using WISE Image Service\footnote{\url{https://irsa.ipac.caltech.edu/applications/wise/}} \citep{2010AJ....140.1868W} with an image cutout of 300 arcsec and we confirmed the absence of nebulosity around these systems. In addition, these systems were not detected as X-ray sources by ROSAT survey\footnote{\url{http://www.xray.mpe.mpg.de/rosat/survey/rass-fsc/}} nor as Gamma-ray sources by Fermi SSC survey\footnote{\url{https://fermi.gsfc.nasa.gov/cgi-bin/ssc/LAT/LATDataQuery.cgi}}. We consider that these Double Periodic Variable stars are optimal targets for further spectropolarimetry studies because these could help to constrain the mechanism based on magnetic dynamo in the donor star proposed by \citet{2017A&A...602A.109S} and these could help us to understand even more about the evolutionary process of DPV stars using models similar to those developed for the interacting binary V495 Centauri by \citet{2019MNRAS.483..862R}.

\ \\

\ \\ 
\footnotesize
\begin{table}[h!]
\caption{New confirmed Double Periodic Variables stars in the southern hemisphere and their respective orbital ($P_{o}$) and long ($P_{l}$) periods. Epochs for the minimum brightness of the orbital light curve and maximum brightness of the long cycle light curve are given and the value $\rm {T^{\star}}= T-2450000$. The brightness values are from ASAS-SN Catalog of Variable stars: II. The Apparent magnitudes were obtained from SIMBAD and APASS-DR9.} 
\begin{center}
\resizebox{14cm}{!}{
$	
\begin{tabular}{|l|l|l|l|}
\hline
ASAS-SN ID						& J163056.92-510307.1 	& J165917.75-350652.9 	& J212958.78-230007.2 	\\
\hline
\hline
Other ID						& GDS J1630570-510306	& V593 Sco				& TYC 6939-678-1 		\\	
RA (hh mm ss)					& 16 30 56.918			& 16 59 17.753			& 21 29 58.778			\\
DEC	(dd mm ss)					& -51 03 07.056			& -35 06 52.884			& -23 00 07.164			\\
\textrm{P$_{o}$ (days)}			& 10.200(1)				& 17.502(8)				& 20.140(4)				\\
\textrm{P$_{l}$ (days)}			& 301.824:				& 582.610:				& 615.733:				\\
T$^{\star}_{0}$(min$_{o}$)		& 7457.85259			& 7670.49845			& 7191.75781			\\
T$^{\star}_{0}$(max$_{l}$)		& 7607.56973			& 7561.72467			& 7675.76706			\\
V (SIMBAD) 						& 12.919(38)*			& 13.523(36)*			& 11.640(140)			\\
B (SIMBAD)		 				& 13.788(38)*			& 14.404(90)*			& 12.240(180)			\\
\hline 
\end{tabular}
$}
\vskip 0.2cm
\scriptsize Note: The apparent magnitudes marked with the asterisk symbol ($\ast$) were obtained from APASS-DR9\footnote{\url{http://vizier.u-strasbg.fr/viz-bin/VizieR?-source=II\%2F336}}.
\end{center}
\label{Table 1}
\end{table}
\normalsize

\vskip 3cm
\section*{Acknowledgements}

We acknowledge support by Fondecyt  1190621.


\begin{thebibliography}{99}

\bibitem[Applegate \& Patterson(1987)]{1987ApJ...322L..99A} Applegate, J.~H., \& Patterson, J.\ 1987, {\it apjl} , {\bf 322}, \href{http://adsabs.harvard.edu/abs/1987ApJ...322L..99A}{L99}

\bibitem[Barr{\'{\i}}a et al.(2013)]{2013A&A...552A..63B} Barr{\'{\i}}a, D., Mennickent, R.~E., Schmidtobreick, L., et al.\ 2013, {\it A\&A}, {\bf 552}, \href{http://adsabs.harvard.edu/abs/2013A\%26A...552A..63B}{A63} 





\bibitem[Garc{\'e}s L et al.(2018)]{2018MNRAS.477L..11G} Garc{\'e}s L, J., Mennickent, R.~E., Djura{\v s}evi{\'c}, G., Poleski, R., \& Soszy{\'n}ski, I.\ 2018, {\it mnras}, {\bf 477}, \href{http://adsabs.harvard.edu/abs/2018MNRAS.477L..11G}{L11} 



\bibitem[Jayasinghe et al.(2019)]{2019MNRAS.486.1907J} Jayasinghe, T., Stanek, K.~Z., Kochanek, C.~S., et al.\ 2019, {\it mnras}, {\bf 486}, \href{https://ui.adsabs.harvard.edu/abs/2019MNRAS.486.1907J}{1907} 


\bibitem[Kochanek et al.(2017)]{2017PASP..129j4502K} Kochanek, C.~S., Shappee, B.~J., Stanek, K.~Z., et al.\ 2017, {\it pasp}, {\bf 129}, \href{https://ui.adsabs.harvard.edu/abs/2017PASP..129j4502K}{104502} 


\bibitem[Mennickent(2017)]{2017SerAJ.194....1M} Mennickent, R.~E.\ 2017, {\it Serbian Astronomical Journal}, {\bf 194}, \href{http://adsabs.harvard.edu/abs/2017SerAJ.194....1M}{1}
\bibitem[Mennickent et al.(2016)]{2016MNRAS.455.1728M} Mennickent, R.~E., Otero, S., \& Ko{\l}aczkowski, Z.\ 2016, {\it mnras}, {\bf 455}, \href{http://adsabs.harvard.edu/abs/2016MNRAS.455.1728M}{1728}
\bibitem[Mennickent et al.(2015)]{2015MNRAS.448.1137M} Mennickent, R.~E., Djura{\v s}evi{\'c}, G., Cabezas, M., et al.\ 2015, {\it mnras}, {\bf 448}, \href{http://adsabs.harvard.edu/abs/2015MNRAS.448.1137M}{1137} 
\bibitem[Mennickent et al.(2012)]{2012MNRAS.427..607M} Mennickent, R.~E., Ko{\l}aczkowski, Z., Djurasevic, G., et al.\ 2012, {\it mnras}, {\bf 427}, \href{http://adsabs.harvard.edu/abs/2012MNRAS.427..607M}{607} 
\bibitem[Mennickent et al.(2008)]{2008MNRAS.389.1605M} Mennickent, R.~E., Ko{\l}aczkowski, Z., Michalska, G., et al.\ 2008, {\it mnras}, {\bf 389}, \href{http://adsabs.harvard.edu/abs/2008MNRAS.389.1605M}{1605}
\bibitem[Mennickent et al.(2003)]{2003A&A...399L..47M} Mennickent, R.~E., Pietrzy{\'n}ski, G., Diaz, M., \& Gieren, W.\ 2003, {\it A\&A}, {\bf 399}, \href{http://adsabs.harvard.edu/abs/2003A\%26A...399L..47M}{L47} 





\bibitem[Poleski et al.(2010)]{2010AcA....60..179P} Poleski, R., Soszy{\'n}ski, I., Udalski, A., et al.\ 2010, {\it AcA}, {\bf 60}, \href{http://adsabs.harvard.edu/abs/2010AcA....60..179P}{179}


\bibitem[Rosales et al.(2019)]{2019MNRAS.483..862R} Rosales, J.~A., Mennickent, R.~E., Schleicher, D.~R.~G., \& Senhadji, A.~A.\ 2019, {\it mnras}, {\bf 483}, \href{http://adsabs.harvard.edu/abs/2019MNRAS.483..862R}{862} 
\bibitem[Rosales Guzm{\'a}n et al.(2018)]{2018MNRAS.476.3039R} Rosales Guzm{\'a}n, J.~A., Mennickent, R.~E., Djura{\v s}evi{\'c}, G., Araya, I., \& Cur{\'e}, M.\ 2018, {\it mnras}, {\bf 476}, \href{http://adsabs.harvard.edu/abs/2018MNRAS.476.3039R}{3039}

\bibitem[Rosales(2018)]{2018IBVS.6248....1R} Rosales, J.~A., Mennickent, R.~E.\ 2018, {\it Information Bulletin on Variable Stars}, {\bf 6248}, \href{http://adsabs.harvard.edu/abs/2018IBVS.6248....1R}{1} 


\bibitem[Shappee et al.(2014)]{2014ApJ...788...48S} Shappee, B.~J., Prieto, J.~L., Grupe, D., et al.\ 2014, {\it apj}, {\bf 788}, \href{https://ui.adsabs.harvard.edu/abs/2014ApJ...788...48S}{48} 


\bibitem[Schleicher \& Mennickent(2017)]{2017A&A...602A.109S} Schleicher, D.~R.~G., \& Mennickent, R.~E.\ 2017,{\it A\&A}, {\bf 602}, \href{http://adsabs.harvard.edu/abs/2017A\%26A...602A.109S}{A109}


\bibitem[Stellingwerf(1978)]{1978ApJ...224..953S} Stellingwerf, R.~F.\ 1978, {\it apj}, {\bf 224}, \href{http://adsabs.harvard.edu/abs/1978ApJ...224..953S}{953} 




\bibitem[Wright et al.(2010)]{2010AJ....140.1868W} Wright, E.~L., Eisenhardt, P.~R.~M., Mainzer, A.~K., et al.\ 2010, {\it aj}, {\bf 140}, \href{http://adsabs.harvard.edu/abs/2010AJ....140.1868W}{1868}



















\end{thebibliography}
\end{document}